\documentstyle[12pt]{article}
\begin{document}
\baselineskip = 24pt

\begin{titlepage}
\vspace{1cm}
\begin{center}{\large {\bf Finite-difference representations of the degenerate
affine Hecke algebra}}\\
\vspace{1.5cm}
D.Uglov \\
\vspace{1 cm}
{\it Department of Physics, State University of New York at Stony Brook} \\
{\it Stony Brook, NY 11794-3800} \\
e-mail: denis@max.physics.sunysb.edu  \\
\vspace{0.5cm}
September 26, 1994

\vspace{3cm}

\begin{abstract}
The representations of the degenerate affine Hecke algebra in which the
analogues of the Dunkl operators are given by finite-difference operators are
introduced. The non-selfadjoint lattice analogues of the spin
Calogero-Sutherland hamiltonians are analysed by Bethe-Ansatz. The $
sl(m)$-Yangian representations arising from the finite-difference
representations of the degenerate affine Hecke algebra are shown to be related
to the Yangian representation of the 1-d Hubbard Model.
\end{abstract}

\end{center}
\end{titlepage}

\section{Introduction}
The Calogero-Sutherland model and its spin generalizations discovered in
\cite{Ber1} have recently come to occupy a prominent place among exactly
solvable 1+1 dimensional many-particle dynamical systems. The chief interest of
these models for solid-state physics lies in the fractional character of their
quasi-particle excitations. On the other hand remarkable relations were found
between these models and the long-range interacting Haldane-Shastry spin chains
and Conformal Field Theory \cite{Ber1,Ber2}.

One of the attractive features of the Calogero-Sutherland model is its relative
simplicity as compared to, for example, the models solvable by QISM such as the
Heisenberg spin chain. In particular the wave functions are sufficiently simple
as to permit explicit computation of some correlation functions
\cite{Correlators,Pasquier}. This simplicity is ultimately due to the close
relationship that exists between these models and the representation of the
degenerate affine Hecke algebra given by the differential Dunkl operators
\cite{Ber1,Dunkl,Polychronakos}. This algebra also gives rise to the Yangian
symmetry in the higher-spin Calogero-Sutherland model and the Haldane-Shastry
spin chain \cite{Ber1}.

The main aim of the present letter is to report upon the finite-difference
representations of the degenerate affine Hecke algebra. In these
representations the analogues of the differential Dunkl operators are
difference operators that are acting on the functions of several integer
variables running through all integer numbers. There are a pair of such
representations - one contains the left finite-differences, another - the right
ones. These representations are referred to as the ``left'' and the ``right''
in what follows. The left and right operators do not commute.  Like in the
continuum case the representations of the degenerate affine Hecke algebra
generate representations of the $sl(m)$-Yangian in the space of wave functions
of the lattice bosons or fermions with $m$ internal degrees of freedom. The
generators of such Yangian representations can in principle be expressed in
terms of the creation and annihilation operators for lattice bosons or
fermions. In the case of the fermions of spin 1/2 the expre
ssions obtained this way are closely related to the fermionic representation of
$sl(2)$-Yangian described in \cite{Hubbard}.

As for the spin Calogero-Sutherland model \cite{Ber1} the quantum determinant
of the Yangian can be used to obtain a family of mutually commuting operators
that also commute with the Yangian. These operators then are  natural
candidates for the discrete counterparts of the Calogero-Sutherland
hamiltonian. The simplest nontrivial operators that one obtains in this fashion
from the left and right underlying representations of the degenerate affine
Hecke algebra are found to be the left- and the right-hopping parts ( denote
them by $ h_L $ and $ h_R $ ) of the higher spin Hubbard hamiltonians ( denote
such a hamiltonian by $ H_H $ ) or of their bosonic analogues:  $ H_H = h_L +
h_R $ ; $( h_L )^{\dagger} = h_R $ . The operators $ h_L , h_R $ are mutually
adjoint and do not commute i.e. they are not normal and cannot be brought to
the diagonal form like the Calogero-Sutherland hamiltonian. Still one can
analyse these operators separately and find their characteristic numbers ( that
are in general complex ) and c
orresponding eigenfunctions. This is done with the aid of the coordinate or
algebraic Bethe-Ansatz. The continuum limit taken in the explicit
eigenfunctions of the operators $ h_L, h_R $ then yields the eigenfunctions of
the continuum Hamiltonians that describe particles with $ \delta $-interaction
\cite{Yang}.

The contents of the letter are as follows: In sec. 2 the definition of the
degenerate affine Hecke algebra is recalled and its finite-difference
representations are introduced. In sec. 3 the Yangian representations arising
from the difference representations of the degenerate affine Hecke algebra are
discussed and compared to the Yangian representation of the Hubbard model. Then
the operators $ h_L $ and $ h_R $ are introduced. Finally these operators are
analysed by algebraic and coordinate Bethe-Ansatz in  sec. 4.

\section{The finite-difference representations of the degenerate affine Hecke
algebra}

The degenerate affine Hecke algebra is an extension of the symmetric group $
S_N $ by $ N $ additional generators $  d_i $ , $1\leq i\leq N $. Denote the $
N-1 $ generators of $ S_N $ by $ K_{ii+1} $ , $ 1\leq i \leq N-1 $. The
defining relations of the degenerate affine Hecke algebra then read
\cite{Ber1,Pasquier}:
\begin{eqnarray}
K_{ii+1}K_{i+1i+2}K_{ii+1} & = & K_{i+1i+2}K_{ii+1}K_{i+1i+2} \: ; \nonumber \\
K_{ii+1}^2 & =  &1 ; \nonumber  \\ \mbox{}
[ K_{ii+1}, d_k ] & =  & 0 \: , \: k \neq i,i+1 \:  ;  \nonumber \\
K_{ii+1}d_i - d_{i+1}K_{ii+1} & = & \mbox{} \lambda  \: ; \nonumber \\ \mbox{}
[ d_i , d_j ] &  = &  0 .
\end{eqnarray}
Where $ \lambda $ is a complex number - parameter of the algebra.

The relations (1) have a well-known representation
\cite{Dunkl,Polychronakos,Ber1} in which the generators act on the functions of
$ N $ variables $ \{z_i\}_{1\leq i \leq N} $. The $ K_{ii+1} $ are represented
by permutation of the arguments:  $  ( K_{ij}f )(z_1, \cdots, z_i,\cdots, z_j,
\cdots, z_N)\; = \; f(z_1,\cdots,z_j,\cdots,z_i,\cdots,z_N) $ ; and the $ d_i $
are represented by the differential operators ( Dunkl operators ):
\begin{equation}
d_i \rightarrow  \partial_{z_i} + \lambda \sum_{j \neq i } \frac{z_i}{z_i -
z_j} K_{ij} - \lambda \sum_{j < i} K_{ij}\; .
\end{equation}

Let now $ f(x_1,\cdots,x_N) $ be a function of variables $ x_i $ , $ 1\leq
i\leq N $ and each of the $ x_i $ runs through all integer numbers. In the
linear space of such functions one has two representations of the degenerate
affine Hecke algebra. Introduce some notations. Let $ {\Delta}_i^{\pm} $ denote
the finite-differences: $ ({\Delta}_i^{\pm}f)(\cdots,x_i,\cdots)\;=\;
f(\cdots,x_i\pm1,\cdots)$, and $ \theta^{\pm}_{ij} $ - the step-functions: $
\theta^{\pm}_{ij}\; \equiv \; \theta^{\pm}(x_i- x_j) $ where:
\[           \theta^+(x)  =  \left\{ \begin{array}{ll}
             1 & \mbox{if $ x\geq0$ , } \\
             0 & \mbox{if $ x<0 $ . }
             \end{array}
        \right.
\theta^-(x)  =  \left\{ \begin{array}{ll}
1 & \mbox{if $ x>0$ , }  \\
0 & \mbox{if $ x\leq0 $ . }
\end{array}
\right.    \]

Let $ K_{ij} $ be as before the permutation of the $i$-th and $j$-th
coordinates . Then the operators:
\begin{equation}
d_{i} = \Delta_i^+ + \lambda \sum_{j\neq i} \theta^+_{ij} K_{ij} - \lambda
\sum_{j<i} K_{ij}
\end{equation}
satisfy the defining relations of the degenerate affine Hecke algebra (1). In
what follows this representation will be referred to as the ``left''
representation. There is another - ``right'' representation in which the $
d_{i} $-s  are given by:
\begin{equation}
d_{i} = \Delta_i^- + \lambda \sum_{j\neq i} \theta^-_{ij} K_{ij} - \lambda
\sum_{j<i} K_{ij}
\end{equation}

The proof that the difference operators (3,4) satisfy the relations (1) is
straightforward and
uses the following easily verified properties of the step-functions of integer
arguments and finite-differences:
\begin{eqnarray*}
\theta_{ij}^{\pm}\theta_{ik}^{\pm}+ \theta_{ik}^{\pm}\theta_{jk}^{\pm}-
\theta_{ij}^{\pm}\theta_{jk}^{\pm} & = & \theta_{ik}^{\pm}\;, \\
(\Delta_i^{\pm}\theta_{ji}^{\pm}) - \theta_{ji}^{\pm}& = & \mp
\delta_{x_i\,x_j}\; , \\
\theta_{ij}^{\pm} + \theta_{ji}^{\pm} & = & 1 \pm \delta_{x_i\,x_j}\;,
\end{eqnarray*}
here $ \delta_{x\,y}$ is the Kronecker delta. Note, that for the second
equation above to hold it is essential that the $ x_i $-s run through all
integer numbers.

\section{The Yangian representations generated by the finite-difference
representations of the affine Hecke algebra}

Once the representations of the degenerate affine Hecke algebra are known it is
straightforward to obtain out of them representations of $sl(m)$-Yangians. The
way to do this for the finite-difference representations is completely
analogous to the continuum case considered in detail in \cite{Ber1}. The
Yangian transfer-matrices then can be in principle expressed in terms of the
lattice bosons or fermions. In the case of the $sl(2)$-Yangian the fermionic
transfer-matrices that one gets from the left and the right representations of
the difference analogs of the Dunkl operators (3,4) yield the fermionic
representation of $Y(sl(2))$ found earlier in the Hubbard model \cite{Hubbard}.
The operators $d_i$ can also be used to derive the ( left or right ) finite
difference analogs of the spin Calogero-Sutherland hamiltonians.

Let now $ f(x_i,s_i;\cdots,;x_N,s_N) $ be a function of integer coordinates
$x_i$ and spins $s_i$, $ 1\leq s_i \leq m $; and $ P_{ij} $ - the spin
permutation operator. The bosonic functions then are characterized by the
condition $ ( K_{ij} - P_{ij} ) f = 0 $ ; and the fermionic - by $ ( K_{ij} +
P_{ij} ) f = 0 $.

As it was shown in \cite{Ber1}, the relations (1)  guarantee that the
transfer-matrices:
\begin{equation}
 T^{\pm}_0(u) = (I \pm \frac{\lambda P_{01}}{u-d_1})(I \pm \frac{\lambda
P_{02}}{u-d_2})\cdots(I \pm \frac{\lambda P_{0N}}{u-d_N})\; ,
\end{equation}
where the subscript $0$ refers to the auxiliary $m$-dimensional vector space
and $u$ is the spectral parameter preserve the space of bosonic $ (+) $ or
fermionic $ (-) $ wave functions. Since the transfer matrices satisfy the
Yang-Baxter relations:
\begin{equation}
(u-v \pm \lambda P_{00^{'}} ) T^{\pm}_0(u)T^{\pm}_{0^{'}}(v) =
T^{\pm}_{0^{'}}(v)T^{\pm}_0(u) (u-v \pm \lambda P_{00^{'}})
\end{equation}
they provide representations of the $sl(m)$-Yangians in the spaces of bosonic $
(+) $ or fermionic $ (-) $ wave-functions. Somewhat freely we call the algebra
(6) a Yangian even though the quantum determinant of $T$ is not assumed to be
equal to one.

{}From the defining relations of the degenerate affine Hecke algebra (1) it
also follows that the polynomial $ C(u)=  (u-d_1)(u-d_2)\cdots(u-d_N) $
commutes with the operators of coordinate permutations $K_{ij}$ and all the
operators $d_1,\cdots, d_N$ \cite{Ber1}. Hence $C(u)$ preserves the spaces of
bosonic and fermionic wave-functions and in each of these spaces it commutes
with the corresponding Yangian transfer-matrix $ T^+ $ or $ T^- $. When the
algebra (1) is represented by the Dunkl operators (2) $ C(u) $ is the
generating function for  the spin Calogero-Sutherland hamiltonians and their
higher-derivative conserved charges \cite{Ber1}. In the case when the $ d_i$-s
are the finite-difference operators (3) or (4) one can consider the
corresponding $C(u)$ as the generating function of the difference analogs of
the Calogero-Sutherland hamiltonians.

The simplest non-trival operator that is given by one of the coefficients of
the polynomial $C(u)$ is just the sum of all $ d_i$-s. Since on the lattice one
has a left-right pair of representations of the degenerate affine Hecke algebra
there are a pair of such operators: $ h_L = \sum_{1\leq i \leq N} d^l $ and $
h_R = \sum_{1\leq i\leq N} d^r $ where $ d^l $ and $ d^r $ mean the operators
given by correspondingly (3) and (4). Explicit forms of the operators $ h_L $
and $ h_R $ look as follows:
\begin{eqnarray}
h_L = \sum_{1\leq i \leq N}\Delta_i^+ + \lambda \sum_{1\leq i<j \leq N}
\delta_{x_i\,x_j} K_{ij}\; , \\
h_R = \sum_{1\leq i \leq N}\Delta_i^- - \lambda \sum_{1\leq i<j \leq N}
\delta_{x_i\,x_j} K_{ij}\; .
\end{eqnarray}
The permutations $ K_{ij} $ appearing in the above expressions after the $
\delta $-symbols will from now on be dropped since  $
(\delta_{x_i\,x_j}K_{ij}f)(x_1,\cdots) = (\delta_{x_i\,x_j}f)(x_1,\cdots) $.

Since the operators $ h_L $ , $ h_R $ preserve the space of the bosonic or
fermionic wave functions one can rewrite them in the secondly quantized form:
\begin{eqnarray}
h_L = \sum_{s\in \mbox{{\bf Z}}} \sum_{\beta=1}^{m} a_s^{\beta
\dag}a_{s+1}^{\beta} + \frac{\lambda}{2}\sum_{s\in \mbox{{\bf Z}}}( n_s^2 -
n_s ) \; ,\\
h_R= \sum_{s\in \mbox{{\bf Z}}} \sum_{\beta=1}^{m} a_{s+1}^{\beta
\dag}a_{s}^{\beta} - \frac{\lambda}{2}\sum_{s\in \mbox{{\bf Z}}}( n_s^2 -  n_s
) \; .
\end{eqnarray}
Here the $ a_{s}^{\beta \dag}\:,\; a_{t}^{\beta} $ are either bosonic or
fermionic creation and annihilation operators on the 1-dimensional lattice: $ [
a_s^{\beta}\:,\; a_t^{\gamma \dag}]_{\mp} = \delta_{st}\delta_{\beta\gamma}\:;
 [ a_s^{\beta \sharp}\:,\; a_t^{\gamma \sharp}]_{\mp} = 0\: ; \; \; s,t \in
\mbox{{\bf Z}}\:, 1\leq \beta,\gamma\leq m $; and $ n_s = \sum_{1\leq \beta
\leq m} a_s^{\beta \dag}a_s^{\beta} $ .

In the subspace with the fixed number of particles $N$ the operators (9,10)
evidently reduce to (7,8). Notice that one can get the ``+'' sign in front of
the coupling constant $ \lambda $ in (10) and (8) by taking the $ h_R $
generated by the right representation of the operators $ d_i $ (4) in which $
\lambda $ is replaced by $ -\lambda $. These $ d_i$-s then will satisfy the
relations (1) with $ \lambda $ replaced by $-\lambda$. Subsequently it is
assumed that the operator denoted $ h_R $ is given by (10) with $ \lambda
\rightarrow -\lambda $. Then if $ H_H $ is the hamiltonian of the Hubbard model
of spin $ (m-1)/2 $ or its bosonic counterpart one has: $ H_H = h_L + h_R $.

The operators (9,10) commute with the transfer-matrices (5) which in principle
can be expressed in terms of the creation/annihilation operators. In particular
if the transfer matrix is represented as the series:
\begin{equation}
 T(u) = I + \lambda \sum_{k \geq 0} \frac{1}{u^{k+1}}t^{(n)}
\end{equation}
then the bosonisation(fermionization) of $ t^{(0)}$ and $ t^{(1)} $ yields the
bosonic or fermionic representations of the Yangian Serre relations
\cite{Drinfeld}. Below this is done for the case of the $sl(2)$-Yangian. This
case corresponds to $m=2$ in the operators (9,10).

The auxiliary linear space being 2-dimensional ( $ m=2 $) introduce the
operators: \[ J^{(0)}_{a}\:,\; J^{(1)}_{a} \; ,\; \;  a=1,2,3\; \]
\begin{equation}
J_a^{(k)} = tr_0( s_{0}^a t^{(k)} ) \; , \;k = 0,1 .
\end{equation}
Where $ s_{0}^a \: \equiv \: \frac{1}{2}\sigma_{0}^a $ are acting in the
auxiliary space as designated by the subindex $0$, and $ \sigma^a $ are the
Pauli matrices.

Explicitely these operators read (here $ \pm $ means that the transfer matrix
in (11) comes from the left(+) or the right(-) representation of the degenerate
affine Hecke algebra; and $ \kappa = 1(-1) $ for the bosonic(fermionic) case):
\begin{eqnarray}
J_{a}^{(0)}& = &\sum_{1 \leq i \leq N} s_i^a \;,  \\
J_{a}^{(1)}& = & \sum_{1 \leq i \leq N} s_i^a \Delta_i^{\pm} - \frac{i \kappa
\lambda}{2} \sum_{1 \leq i\neq j \leq N} \varepsilon(x_i-x_j)
\epsilon_{abc}s_i^b\,s_j^c\; \pm  \nonumber \\
 & & \frac{i \kappa \lambda}{2} \sum_{1 \leq i\neq j \leq N}
\delta_{x_ix_j}s_i^a \; + \; \frac{i \kappa \lambda}{2}(N-1) J_{a}^{(0)}\: .
\end{eqnarray}
In these expressions it is understood that the operators $ s_i^a $ act upon
i-th the spin variable of the wave-functions and the notation $ \varepsilon(x)
$ is introduced for the step function: $ \varepsilon(x) = 1(-1)$, if $x>(<)0$;
and $ \varepsilon(0) = 0 $.

It follows from the equation (6) that the operators (13,14) satisfy the
$sl(2)$-Yangian Serre relations \cite{Drinfeld}:
\begin{eqnarray}
[J_{a}^{(0)},J_{b}^{(0)}]& = &i \epsilon_{abc}J_{c}^{(0)}\; \nonumber \\
\mbox{}
[J_{a}^{(0)},J_{b}^{(1)}]& = &i \epsilon_{abc}J_{c}^{(1)}\; \nonumber \\  \mbox
[J_{a}^{(2)},J_{b}^{(1)}] + [J_{b}^{(2)},J_{a}^{(1)}]& = &i\lambda^2
(\epsilon_{acd}\{J_{b}^{(0)},J_{c}^{(0)},J_{d}^{(1)}\}+\epsilon_{bcd}\{J_{a}^{(0)},J_{c}^{(0)},J_{d}^{(1)}\}).
\end{eqnarray}
Where \[ J_{a}^{(2)}=-\frac{i}{2} \epsilon_{abc}[J_{b}^{(1)},J_{c}^{(1)}] \],
and $ \{.,.,.\} $ stands for symmetrization. Notice that the last term in the
right-hand side of (14) can be dropped without affecting the Serre relations,
this is done below in the eq. (17) .

Finally the operators  $ J^{(0)}_{a}\:,\; J^{(1)}_{a} $,$\; a=1,2,3\; $ can be
expressed in terms of the creation/annihilation operators (either bosonic or
fermionic). As before there are two representations (left/right) which are
distinguished by the superscripts $ +/- $ over the generators:
\begin{eqnarray}
J_{a}^{(0)\pm}& = &\sum_{t \in \mbox{{\bf Z}}} a^{\beta
\dag}_t(s^a)_{\beta\,\gamma}a^{\gamma}_t \;,  \\
J_a^{(1)+}& = &\sum_{t\in \mbox{ {\bf Z}} } a^{\beta
\dag}_t(s^a)_{\beta\,\gamma}a^{\gamma}_{t+1} - \frac{i\kappa \lambda}{2}\sum_{r
\neq t} \varepsilon (r-t) \epsilon_{abc}  a^{\beta \dag}_r
(s^b)_{\beta\gamma}a^{\gamma}_r a^{\mu \dag}_t (s^c)_{\mu\nu}a^{\nu}_t  +
\nonumber \\
& & + \frac{\kappa \lambda}{2} \sum_{r \in \mbox{{\bf Z}}}(n_r - 1) a^{\beta
\dag}_r(s^a)_{\beta \gamma}a^{\gamma}_r \; , \nonumber \\
J_{a}^{(1)-}& = &\sum_{t \in \mbox{{\bf Z}}} a^{\beta
\dag}_t(s^a)_{\beta\gamma}a^{\gamma}_{t+1} - \frac{i\kappa \lambda}{2}\sum_{r
\neq t} \varepsilon(r-t) \epsilon_{abc}  a^{\beta
\dag}_r(s^b)_{\beta\,\gamma}a^{\gamma}_r a^{\mu\dag}_t(s^c)_{\mu\nu}a^{\nu}_t -
\nonumber \\
& & - \frac{\kappa \lambda}{2}\sum_{r \in \mbox{{\bf Z}}}(n_r - 1) a^{\beta
\dag}_r(s^a)_{\beta\,\gamma}a^{\gamma}_r \; .
\end{eqnarray}
The $ \kappa $ above refers to the bosonic ($ \kappa = 1 $) or fermionic ($
\kappa = -1 $) cases. The operators $ J^{(0,1)+}_{a} $ commute with the
operator $ h_L $ (9) and $ J^{(0,1)-}_{a} $ commute with the operator $ h_R $
for $ m=2 $.

It is a special feature of the case of the fermions of the spin-1/2, that is
the case relevant to the Hubbard model, that the operators  $ {\bf
J}^{(0)}_{a}\:,\; {\bf J}^{(1)}_{a} $  given by:\begin{eqnarray*}
 {\bf J}^{(0)}_{a} & = & J^{(0)\pm}_{a}\;, \\
{\bf J}^{(1)}_{a} & = & J^{(1)+}_{a} - J^{(1)-}_{a}|_{\lambda \rightarrow
-\lambda} .
\end{eqnarray*}
still satisfy the Serre relations (15) with the $\lambda$ changed to
$2\lambda$. These operators coincide with the generators of the ``spin''
Yangian in the Hubbard model \cite{Hubbard}.

\section{Characteristic numbers and eigenvectors of the operators $ h_L\;,\;h_R
$}
Since the finite-difference operators $ h_L\:,\; h_R $ are not normal with
respect to the usual inner product in the Fock space, one cannot diagonalize
them. Nevertheless the characteristic numbers of these operators and
corresponding eigenvectors can be found by Bethe-Ansatz. Such eigenvectors
apparently are the `` lowest vectors `` in Jordan chains \cite{Gokhberg}
corresponding to the characteristic numbers.

First, consider the pair of operators $ h_L $ and $ h_R $:
\begin{equation}
h_L = \sum_{1\leq s \leq M} \sum_{\beta=1}^{m} a_s^{\beta \dag}a_{s+1}^{\beta}
+ \frac{\lambda}{2}\sum_{1\leq s \leq M}( n_s^2 -  n_s ) \; ,\;\; \; h_R =
h_L^{\dag}
\end{equation}
as defined on a periodic chain of length $M$: $ M+1 \equiv 1 $ . Recall, that
the creation/annihilation operators here can be either bosonic or fermionic.
Besides the hermitian conjugation, the operators $ h_L\:,\;h_R $ are related by
the spatial reflexion: $ \Pi : a_s^{\beta \sharp} \rightarrow a_{M-s+1}^{\beta
\sharp} $; $ h_L = \Pi h_R \Pi $. In this setting the eigenvalues and the
eigenvectors of the operators (18) can be constructed by means of the Algebraic
Bethe-Ansatz. The basic ingredients of the Algebraic Bethe-Ansatz for the
operators (18) i.e. the $R$-matrices, $L$-operators and the trace identities
will now be described.

Due to the great similarity between the bosonic and fermionic cases they are
considered in parallel. Both bosonic and fermionic local $ L $-operators
related to the site with the number $s$ are $ (1+m)\times(1+m) $ matrices with
operator coefficients expressed in terms of the creation/annihilation operators
$ a_{s}^{\beta \sharp}\;,\; 1 \leq \beta \leq m $  :
\begin{equation}
L(u)_s=\left( \begin{array}{cc}
                   u+\lambda\,n_s &  (a_s)_m  \\
                   \kappa\lambda\,(a_s)_m^{\dag} &  I_{m\times m}
                  \end{array}
            \right)
\end{equation}
Where $ (a_s)_m $ is the row: $ ( a_s^1,\cdots, a_s^m ) $, $ I_{m\times m} $ -
$m \times m$ identity matrix, $u$ -the spectral parameter and $ \kappa\;=\;+(-)
$ for bosonic(fermionic) case. For the fermions the matrix (19) is graded: if
$L = (L_{ab})_{1\leq a,b \leq 1+m}$ then $ p(a) = 0 \mbox{, if $a=1$}$; $
p(a)=1 \mbox{, if $ 2\leq a\leq 1+m$} $.

The $L$-operators (19) satisfy the Yang-Baxter relation with the rational
$R$-matrix \cite{R} which is $ sl(1+m) $-symmetric for the bosons and $
sl(1|m)$-symmetric for the fermions :
\begin{equation}
(\lambda + (u-v) P)L(u)_s\otimes L(v)_s = L(v)_s\otimes L(u)_s(\lambda + (u-v)
P)
\end{equation}
$P$ here is the permutation(graded permutation) operator  in $ \mbox{{\bf
C}}^{1+m} \otimes \mbox{{\bf C}}^{1+m}(\mbox{{\bf C}}^{1|m} \otimes \mbox{{\bf
C}}^{1|m})$ and $ \otimes $ means tensor(graded tensor) product for
correspondingly the bosonic(fermionic) case.

The $L$-operators are used to construct the transfer-matrices $\tau(u)$:
\begin{equation}
\tau_{Bose}(u) = tr(L_M(u)L_{M-1}(u)\cdots L_1(u))\:,\;\;;\;\; \tau_{Fermi}(u)
= str(L_M(u)L_{M-1}(u)\cdots L_1(u))\:,\;
\end{equation}
And the operators $ h_L $ are then obtained from the transfer-matrices by means
of the trace identities:


\begin{equation}
\tau(u) = u^M + u^{M-1}\lambda N + u^{M-2}\lambda (h_L - \frac{\lambda}{2}N +
\frac{\lambda}{2}N^2) + {\cal O}(u^{M-3}),
\end{equation}
where $N$ is the particle number. The trace identities are the same in form for
the bosons and for the fermions.

The operators $h_R = h_L^{\dag}$ are obtained in the same way from the
transfer-matrices $ \tau({\overline u})^{\dag} = \Pi \tau(u)\Pi =
(s)tr(L_1(u)L_2(u),\cdots,L_M(u))$.

Since the $ L $-operators (19) become lower-triangular when applied to the
local Fock pseudovacuum, the well-known machinery of the nested Algebraic Bethe
Ansatz \cite{NestedBA} can be employed to construct a family of eigenvectors of
the operators $ h_L\:,\; h_R $ and the corresponding eigenvalues. The structure
of these eigenfunctions and eigenvalues, however, is more transparent if one
uses the coordinate Bethe-Ansatz approach \cite{Bethe}.

The coordinate Bethe-Ansatz gives the family of eigenfunctions $f(x_1,\cdots,
x_N|\{z_i\})$ of the operator $h_L$ (7) in the form of Bethe sums parametrized
by $ N $ complex numbers $ z_1,\cdots, z_N $ : $ h_Lf(\{z_i\}) = (\sum_{1\leq
i\leq N}\,z_i)\, f(\{z_i\})$ . In the sector $ x_1 \leq x_2 \leq \cdots \leq
x_N $ the explicit expression for the eigenfunction is:
\begin{eqnarray}
f(x_1,\cdots,x_N|\{z_i\})& = & \sum_{R\in S_N} \phi(R)
z_{R_1}^{x1}z_{R_2}^{x2}\cdots z_{R_N}^{xN} \:, \\
\phi((s\leftrightarrow s+1)R)&=&-\frac{\lambda - \kappa (z_{R_s} -
z_{R_{s+1}})P_{s\,s+1}}{z_{R_s} - z_{R_{s+1}} + \lambda}\phi(R)
\end{eqnarray}

Where it is understood that the $f$ and $\phi$ depend also on $N$ spins and the
$f$ is either bosonic ($ \kappa = 1 $ ) or fermionic ( $ \kappa = -1 $ ). On
the infinite lattice the above expressions provide eigenfunctions of $ h_L $ in
the algebraic sense. If lattice is periodic than as usual the restrictions on
the set of $ z_i $ and the functions of spins $ \phi(R) $ arise from the
periodic boundary conditions.

The eigenfunctions (23) resemble very much the exact eigenfunctions for the
continuum particles with $ \delta $ -interaction \cite{Yang} and converge to
them in the continuum limit:
\begin{equation}
z_i = e^{i\epsilon k_i}\;, x_i = \mbox{ x}_i/\epsilon \;, \lambda = \epsilon
c\;\; , \epsilon \rightarrow 0 .
\end{equation}
Here $ k_i $ are the set of momenta, $ \mbox{x}_i $ - continuous coordinates of
the particles and  $c$ - the coupling constant for the continuum particles with
$ \delta $ -interaction.

\section{Conclusions}
In this letter we described the finite-difference representations of the affine
Hecke algebra and the lattice analogues of the spin Calogero-Sutherland
hamiltonians. These lattice operators are not selfadjoint yet their explicit
eigenfunctions, which can be constructed by means of Bethe-Ansatz, go over in
the continuum limit into the eigenfunctions of the continuum particles with $
\delta $-interaction. The Yangian representations that are generated by the
finite-difference representations of the degenerate affine Hecke algebra were
seen to produce the Yangian representation of the 1-d Hubbard model. The
structure of the Yangian generators in the lattice analogues of the spin
Calogero-Sutherland hamiltonians is very close to that one in the Hubbard model
yet unlike the former models the Hubbard model is not known to be solvable by
Algebraic Bethe-Ansatz, which is detrimental to the understanding of its
correlation functions. It would be interesting then to see if the relation
between the Yangians mentioned a
bove cannot be used to develop Algebraic Bethe-Ansatz for the Hubbard model.

{\bf Acknowledgements} \\
I am grateful to F.H.L.Essler and V.E.Korepin for stimulating discussions. This
work was partially supported by the NSF grant 9309888.


\begin{thebibliography}{99}
\bibitem{Ber1}
D.Bernard, M.Gaudin, F.D.M.Haldane and V.Pasquier, J.Phys.A 26 (1993) 5219
\bibitem{Ber2}
D.Bernard, V.Pasquier and D.Serban, ``Spinons in Conformal Field Theory''
preprint SPhT/94/039 (hep-th-9404050)
\bibitem{Bethe}
H.Bethe, Z.Phys. 71 (1931) 205
\bibitem{Drinfeld}
V.G.Drinfel'd, in: Proc. ICM (Berkeley Univ. Press, Berkeley, CA, 1986);
A.LeClair and F.A.Smirnov, Int.J.Mod.Phys.A 7 (1992) 2997
\bibitem{Dunkl}
C.F.Dunkl, Trans. Am. Soc. 311 (1989) 167
\bibitem{Gokhberg}
I.Gokhberg, P.Lancaster, L.Rodman, ``Invariant subspaces of matrices with
applications'', New York: Wiley 1986
\bibitem{NestedBA}
P.Kulish, Dokl.Akad.Nauk USSR 255 (1980) 323; O.Babelon, H.J.de Vega, C.M.
Viallet , Nucl. Phys. B200 (FS4) (1982) 266
\bibitem{R}
P.Kulish, E.Sklyanin, J.Sov.Math. 19 (1982) 1596
\bibitem{Correlators}
F.Lesage, V.Pasquier and D.Serban, ``Dynamical correlation functions in the
Calogero-Sutherland model'' Saclay preprint, April-94 (hep-th-9405008)
\bibitem{Pasquier}
V.Pasquier, ``A lecture on the Calogero-Sutherland models'' preprint
SPhT/94-060 (hep-th-9405104)
\bibitem{Polychronakos}
A.P.Polychronakos, Phys.Rev.Lett. 69 (1992) 703
\bibitem{Hubbard}
D.Uglov and V.Korepin, Phys.Lett.A 190 (1994) 238
\bibitem{Yang}
C.N. Yang, Phys.Rev.Lett. 20 (1968) 1312
\end{thebibliography}
\end{document}